\documentclass[aps,pra,reprint,groupedaddress,twocolumn]{revtex4-2}
\usepackage{graphicx}
\usepackage{xcolor}
\usepackage{amsmath}
\usepackage{mathtools}
\usepackage{amssymb}
\usepackage{dsfont}
\usepackage{amsthm}
\usepackage{hyperref}

\usepackage{svg}

\begin{document}
\author{Giuseppe Ortolano$^{1,2}$}
\author{Stefano Pirandola$^3$}
\author{Leonardo Banchi$^{1,2}$}
\affiliation{$^1$ Dipartimento di Fisica e Astronomia, Università di Firenze, Via G. Sansone 1, I-50019 Sesto Fiorentino (FI), Italy}
\affiliation{$^2$ Istituto Nazionale di Fisica Nucleare, Sezione di Firenze, via G. Sansone 1, I-50019 Sesto Fiorentino (FI), Italy}
\affiliation{$^3$ Department of Computer Science, University of York, York YO10 5GH, United Kingdom}

\title{Lower Bounding the Secret Key Capacity \\ of Bosonic Gaussian Channels via Optimal Gaussian Measurements}
\begin{abstract}
  We find the maximum rate achievable in the private communication over a bosonic quantum channel with a fully Gaussian protocol based on optimal single-mode Gaussian measurements. This rate establishes a lower bound on the secret rate capacity of the channel. We focus on the class of phase-insensitive Gaussian channels. For the thermal-loss and thermal amplification channels, our results demonstrate the optimality, within the constraints of our analysis, of previously proposed protocols, while also providing a significantly simplified formula for their performance evaluation. For the added noise channel, our rate provides a better lower bound than any previously known.
\end{abstract}
\maketitle
\section*{Introduction}
Quantum key distribution (QKD)~\cite{Bennett1984,Ekert1991,Grosshans2002,Grosshans2003,Gisin2002,Pirandola2020} is a central pillar of quantum information, aimed at establishing provably secure communication. A central task in QKD is to find the secret rate capacity, the rate at which two parties can generate a shared private key, by communicating over a quantum channel. This task is particularly relevant for bosonic quantum channels, which model a wide range of physical interactions in continuous variable (CV) quantum communication~\cite{Braunstein2005,Weedbrook_2012,Serafini2017}.
Among these, Gaussian quantum channels are especially important, as they model practical systems, such as optical fibers and free-space communication. 

In this context, Ref.~\cite{Pirandola_2017} established a general upper bound on the secret rate capacity. For the specific case of pure-loss channels, this bound matches the previously-known lower bound~\cite{Pirandola_2009}, yielding an exact, closed-form expression for the capacity. On the other hand, the capacity for the similarly important thermal-loss, thermal amplification and added noise channels, remains unknown, with separate upper and lower bounds. For these channels, the use of coherent~\cite{Schumacher_1996, Lloyd_1997} and reverse coherent ~\cite{Pirandola_2009, Garcia_2009} information, together with the relative entropy of entanglement~\cite{Vedral1997,Vedral1998, Plenio2001} fails to close the gap.

In the present work, we further the effort in ``closing the gap'' between bounds for the aforementioned phase-insensitive channels by providing a lower bound via the computation of the maximum secret rate achievable by a fully Gaussian protocol operating optimized single-mode Gaussian measurements. Our work confirms previous results for the thermal-loss and thermal-amplification channels, while providing a tighter lower bound for the added-noise channel.

The paper is structured as follows. In Section I, we define the secret rate capacity in the context of our protocol. In Section II, we characterize the capacity for a fully Gaussian protocol, and we present the results for phase-insensitive channels. In Section III, we summarize the conclusions of our work.

\section{Secret key rate}
\begin{figure}[]	\includegraphics[width=0.8\columnwidth]{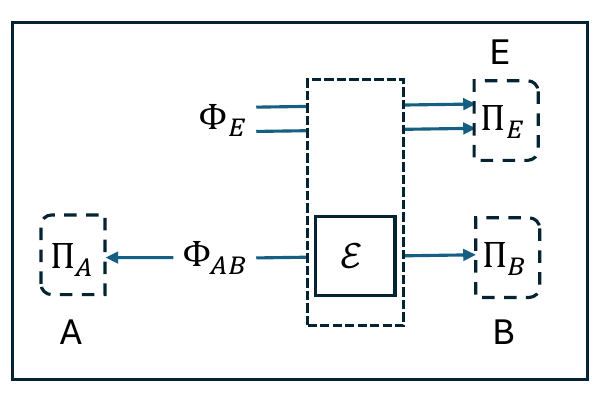}
	\caption{\textit{Communication scheme in the entanglement-based representation.} Alice sends one mode, $B$, of $\Phi_{AB}$ to Bob through a channel $\mathcal{E}$, while retaining mode $A$. Information is exchanged through measurements of the state, and guessing the outcome of either by Alice (direct reconciliation) or Bob (reverse reconciliation). The action of the channel can be interpreted as an eavesdropper, Eve, acting on an extended space to recover shared information. }\label{fig:scheme}
\end{figure}
Consider a communication scheme in the entanglement-based representation that involves two parties, Alice and Bob, as depicted in Fig.~\ref{fig:scheme}.  Alice retains one system (A) of a Two Mode Squeezed Vacuum (TMSV) state, $\Phi_{AB}$,  with parameter $\mu=2 \bar{n}+1$, where $\bar{n}$ is the mean number of photons in each marginal state. In the following we will consider all the relevant quantities in the limit of $\mu\to\infty$. The other system (B) is sent to Bob via a Gaussian channel $\mathcal{E}$, with parameters $\eta$ (transmissivity or gain) and $\omega$ (thermal-noise variance)~\cite{Pirandola_2009}. Communication is achieved by performing a local, single-mode POVM  $\Pi_X= \{\Pi_x\}$ either on A (direct reconciliation) or B (reverse reconciliation). Let us denote in the following the party that performs the encoding measurement as the sender $X=A/B$ and the other party as $Y=B/A$. The rate of communication of this scheme, $R_{XY}$ is upper bounded by the $\chi$-quantity:
\begin{equation}
R_{XY}\leq\chi_{Y}^{\Pi_X}:=S(\rho_{Y})-\sum_x p_x S(\rho_{Y|x}), \label{eq:chiA}
\end{equation}
where $p_x$ is the probability of outcome $x$, $X=A, Y=B$($X=B, Y=A$) in direct (reverse) reconciliation and $\rho_{Y}$ is the total local state of $Y$ and $\rho_{Y|x}$ is the conditioned state of $Y$ to outcome $x$ on system $X$.  In the limit of many uses of the channel and unlimited one-way classical communication the right hand side of Eq.~\eqref{eq:chiA}
is exactly the common randomness between parties $X$ and $Y$ distillable from the classical-quantum system, resulting from the measurement performed on X~\cite{Devetak_2004} (see also Appendix). This means that in this scenario the inequality is indeed saturated, i.e., we have
\begin{equation}
R_{XY}=\chi_{Y}^{\Pi_X}.    
\end{equation}

The action of a possible eavesdropper, Eve, can be analyzed by a dilation of the channel. For Gaussian channels the collective attack performed by Eve can be characterized using an input TMSV on $E$, $\Phi_E$, that defines an unique dilation for the channel, $\tilde{\mathcal{E}}$, up to local unitary transformations~\cite{Pirandola_2008}. The scheme is reported in Fig.~\ref{fig:scheme}. The mutual information of $E$ and the sender system $X$ is upper bounded by:
\begin{equation}
I(E:X)\leq\chi_{E}^{\Pi_X}:=S(\rho_{E})-\sum_x p_x S(\rho_{E|x}). \label{eq:chiE}
\end{equation}

In view of the achievability of the rate in Eq.(\ref{eq:chiA}) (see Appendix), the rate $R$ of private communication through the channel, that establishes a lower bound on the secret rate capacity $K$, is lower bounded as follows:
\begin{equation}
K\geq R\geq \mathcal{L}:=\lim_{\mu\to\infty}  \max_{\Pi_X} \chi_{Y}^{\Pi_X}-\chi_{E}^{\Pi_X}, \label{eq:rate} 
\end{equation}
where for clarity we explicitly stated that the quantities are evaluated in the limit $\mu\to\infty$, and we will omit this limit hereinafter for brevity of notation.  

The total local states $\rho_X$ and $\rho_E$ are not changed by the sender measurement, so the bound $\mathcal{L}$ can be written as:
\begin{align}
&\mathcal{L}=I^{C/RC}_{\mathcal{E}} +\Delta, \\
\Delta&:=\max_{\Pi_X} \sum_x p_x \left(S(\rho_{E|x})-S(\rho_{Y|x})\right), \nonumber \label{eq:bound}
\end{align}
where $I^{C/RC}_{\mathcal{E}}=S(\rho_{A/B})-S(\rho_{AB})$ is the direct/reverse coherent information of the channel~\cite{Pirandola_2017} , and we used the fact that the total state of systems $ABE$ is pure, so that $S(\rho_{E})=S(\rho_{AB})$.
Note how $I^{C/RC}_{\mathcal{E}}$ in itself represents a lower bound for the private capacity of the channel~\cite{Pirandola_2017}, and we will use it as a comparison for our lower bound $\mathcal{L}$, that will then be an improvement on the former if $\Delta>0$ and $\mathcal{L}>0$.
\section{Gaussian measurements}
In the following we restrict the maximization in Eq.(\ref{eq:rate}) to Gaussian measurements and we denote the restricted quantities $\mathcal{L}^G$ and $\Delta^G$. For these measurements the conditional covariance matrix (CM) of the states, $\rho_{Y/E|x}$ will not depend on the measurement outcome $x$~\cite{Weedbrook_2012,Pirandola_2014}, $ S(\rho_{Y/E|x})=S(\rho_{Y/E|X})$ $\forall x$, thus:
\begin{equation}
\Delta^G=\max_{\Pi_X^G} S(\rho_{E|X})-S(\rho_{Y|X}). \label{eq:deltaG}
\end{equation} 

Let us denote the covariance matrix (CM) of the Gaussian state shared by Alice and Bob (after the channel) as:
\begin{equation}
	\mathbf{V}_{YX}=\begin{pmatrix}
		\mathbf{Y}& \mathbf{C} \\
		\mathbf{C}^T & \mathbf{X}
	\end{pmatrix}.
\end{equation}
A local Gaussian measurement on X can be represented as a projection on a Gaussian state $\rho_0$ having mean $\bf{x}_0=0$ and CM $\bf{V}_0$ followed by a displacement of $\mathbf{k}$, indicating the outcome of the measurement~\cite{Weedbrook_2012,Pirandola_2014}. The state of Y conditioned to X measuring $\mathbf{k}$ is Gaussian with CM~\cite{Weedbrook_2012}:
\begin{equation}
	\mathbf{V}_{Y|X}=\mathbf{Y}-\mathbf{C}(\mathbf{X}+\mathbf{V}_0)^{-1}\mathbf{C}^T.  \label{eq:CmC}
\end{equation}
Furthermore we can always decompose the single mode CM $\mathbf{V}_0$ as~\cite{Weedbrook_2012}:
\begin{equation}
	\mathbf{V}_0(\gamma,r,\theta)=\gamma\mathbf{R}(\theta)\mathbf{S}(2r)\mathbf{R}(\theta)^T, 
\end{equation}
where $\gamma=2n_\gamma+1$ is the parameter of a thermal state with $n_\gamma$ photons, $\mathbf{R}$ a rotation of $\theta$ and $\mathbf{S}$ a squeezing with parameter $2r$.

The optimal attack performed by Eve depends on the channel considered, as we detail in the following. Regardless of the specific interaction, we denote the off-diagonal elements of the CM of E and X after the attack, $\mathbf{V}_{EX}$, as $\mathbf{C}_E$. In view of Eq.(\ref{eq:CmC}), Eve conditional CM is 	$\mathbf{V}_{E|X}=\mathbf{E}-\mathbf{C}_E(\mathbf{X}+\mathbf{V}_0)^{-1}\mathbf{C}_E^T $.

The entropy of a Gaussian state is completely determined by its CM, so from the analysis above, it follows that the maximization over the measurement $\Pi_X^G$ of Eq.(\ref{eq:deltaG}) can be performed over the real parameters $(\gamma,r,\theta)$ characterizing $\mathbf{V}_0$.

In the following we compute $\Delta^{G}$, and consequently the lower bound $\mathcal{L}^G            $, for the class of phase insensitive Gaussian channels, composed of thermal loss, thermal amplification and added noise. For these channels we show in Appendix that the conditioned entropies do not depend on the rotation angle $\theta$, nor on the squeezing parameter $r$, so that $\Delta_G$ simplifies as:
\begin{equation}
\Delta^G=\max_{\gamma\geq1} S(\rho_{E|X})-S(\rho_{Y|X}), \label{eq:deltaPI}
\end{equation}
with the maximization being only on the single real parameter $\gamma\geq1$.

Note how the bound $\mathcal{L}^G$ applies either in direct or reverse reconciliation regardless of the channel, so both are computed, but we present only the one that is relevant in each case.
\begin{figure}[]	\includegraphics[width=\columnwidth]{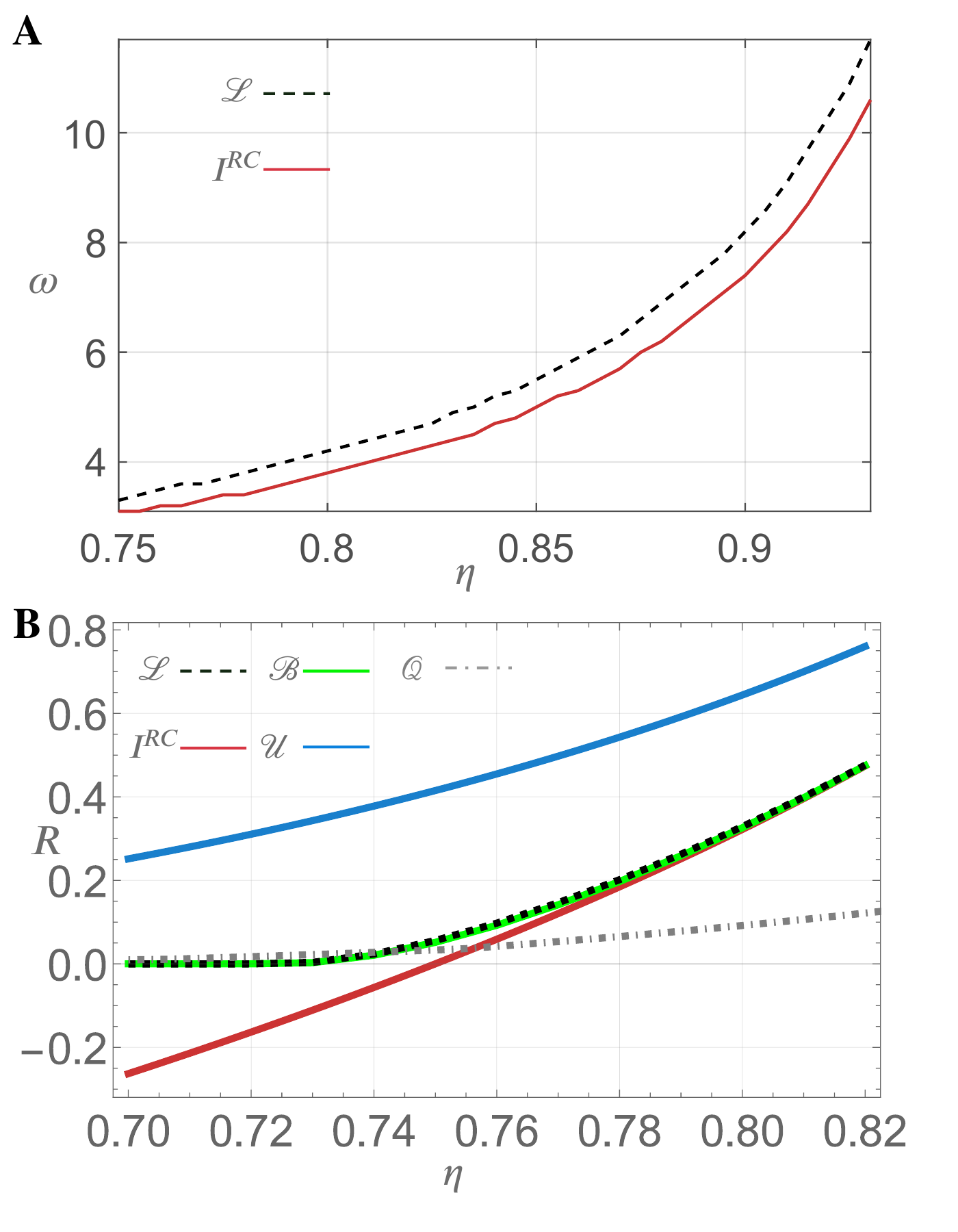}
	\caption{\textit{Thermal-Loss channel}. \textbf{A}. We plot the security threshold $\omega_{th}$, defined in the main text, as a function of the transmissivity $\eta$, for the bound $\mathcal{L}_{\omega_\eta}$ and the channel's reverse coherent information $I^{RC}_{\eta,\omega}$. \textbf{B}. $\mathcal{L}_{\omega,\eta}$ is compared with the lower bounds $\mathcal{B}_{\omega,\eta}$  and $\mathcal{Q}_{\omega,\eta}$ (see the main text) at a fixed value $\omega=3$. We also plot $I^{RC}_{\eta,\omega}$ as well as the upper bound on the capacity $\mathcal{U}_{\omega,\eta}$.}\label{fig:Fig1}
\end{figure}
\section*{Thermal loss}
A thermal loss channel, $\mathcal{E}_{\eta,\omega}$, is a Gaussian channel characterized by parameters $0<\eta<1$, representing the transmissivity and $\omega=2 n_{th} + 1 >1 $, where $n_{th}>0$ is the average number of thermal photons. This channel acts on a single mode Gaussian state of mean $\bf{x}$ and CM $\bf{V}$ with the following input-output relations:
\begin{align}
	\bf{x}\to\eta\bf{x}, && \bf{V}\to \eta\bf{V}+ (1-\eta)\omega \mathds{1}.
\end{align}
The maximum rate in this case is obtained by a reverse reconciliation protocol. The reverse coherent information is~\cite{Pirandola_2009}:
\begin{equation}
	I^{RC}_{\eta,\omega}= -\log_2(1-\eta)-h(\omega).
\end{equation}

 Eve's optimal attack is a collective attack, that for the thermal loss channel consists of an entanglement cloner, detailed in the appendix. The conditional entropies $S(\rho_{A/E|x})$ can be computed from the CMs $	\mathbf{V}_{Y/E|X}$, to yield:
\begin{equation}
\Delta^G_{\eta,\omega}=\max_{\gamma>1} h(\lambda_+^{(E)})+h(\lambda_-^{(E)})-h\left(\frac{\gamma+(1-\eta)\omega}{\eta}\right), \label{eq:deltaLoss eta}
\end{equation}
where $h(x)=\frac{x+1}{2}\log_2[(x+1)/2]-\frac{x-1}{2}\log_2[(x-1)/2]$ is the thermal entropy function and $\lambda_\pm^{(E)}$ are the sympletic eigenvalues of $\mathbf{V}_{E|X}$ (see Appendix for detailed expressions).
The maximization over $\gamma$ is evaluated numerically. Let us denote the maximum amount of noise $\omega_{th}(\eta)$ such that $\mathcal{L}^G_{\eta,\omega}>0$, as the security threshold.  In Fig.~\ref{fig:Fig1}.\textbf{A} we plot $\omega_{th}(\eta)$ as a function of the transmissivity $\eta$ and compare it to the security threshold of $I^{RC}_{\mathcal{E}_{\eta,\omega}}$, showing that $\mathcal{L}^G_{\eta,\omega}$ is indeed a tighter bound than the former. In Fig.~\ref{fig:Fig1}.\textbf{B} we plot the bound as a function of $\eta$ at a fixed value of $\omega$. We compare $\mathcal{L}^G_{\eta,\omega}$ with $I^{RC}_{\mathcal{E}_{\eta,\omega}}$, as well as with the bound of Ref.~\cite{Ottaviani_2016}, that we denote as $\mathcal{B}_{\eta,\omega}$, and the one from Ref.~\cite{Mele_2025}, denoted $\mathcal{Q}_{\eta,\omega}$ that, to the best of our knowledge, jointly make up the best lower bound up to date on the secret key capacity of the thermal loss channel. We also plot the upper bound on the capacity, $\mathcal{U}_{\eta,\omega}=-\log_2(\left(1-\eta)^{n_{th}}\right)-h(\omega)$~\cite{Pirandola_2017}, as a reference. Our lower bound, $\mathcal{L}^G_{\eta,\omega}$, improves on the reversed coherent information, as expected. Notably, the protocol achieving $\mathcal{B}_{\eta,\omega}$ is a special case of our general protocol. Our result show that the former achieves the optimal performance for a fully Gaussian protocol with single mode measurements. On the other hand, $\mathcal{Q}_{\eta,\omega}$  still provides a slightly better bound for lower values of $\eta$.
\begin{figure}[]	\includegraphics[width=\columnwidth]{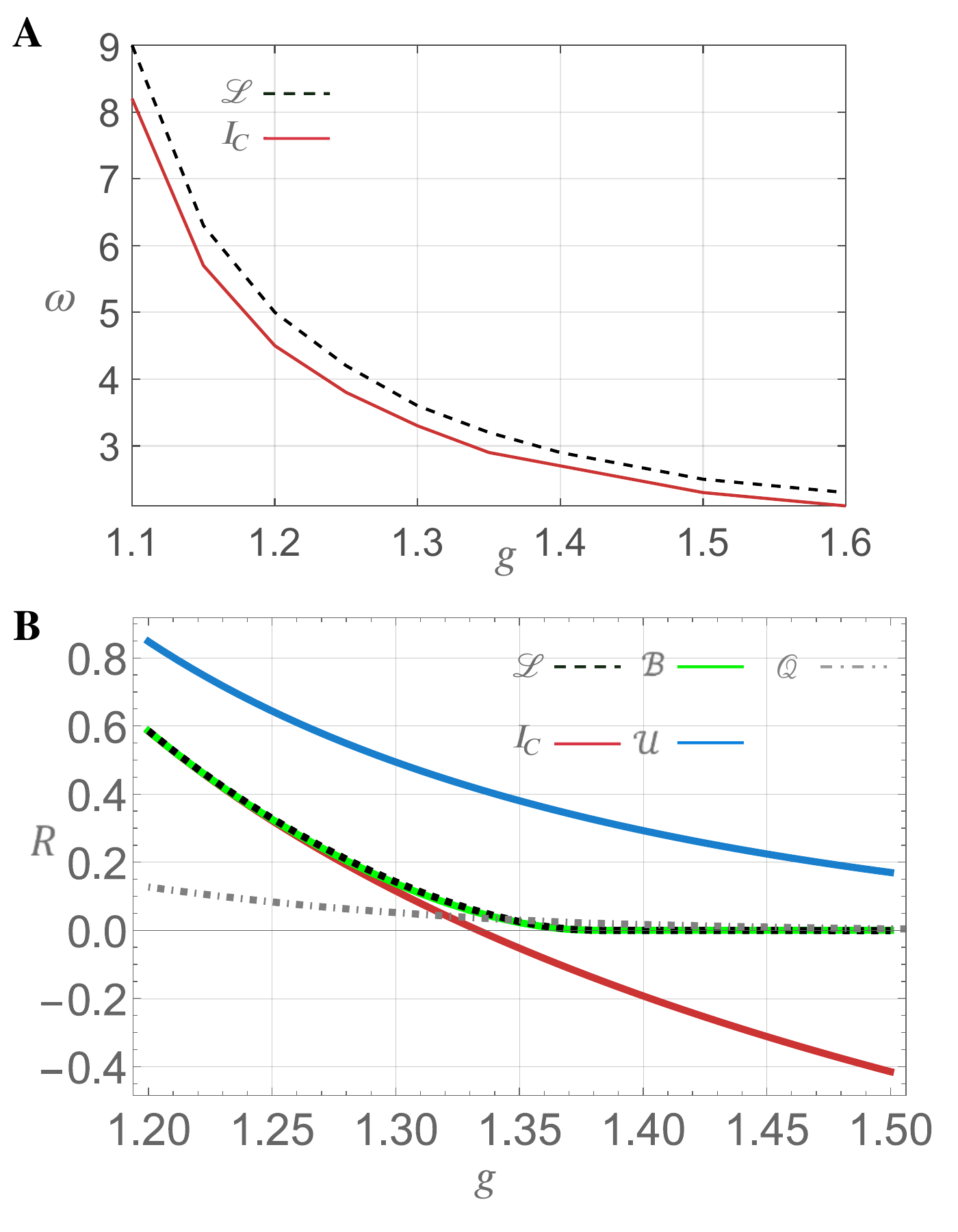}
		\caption{\textit{Thermal Amplification}. \textbf{A}. Security threshold $\omega_{th}$, for the thermal amplification channel (see main text) as a function of the gain $g$, for the bound $\mathcal{L}_{\omega,g}$ and the reverse coherent information of the channel $I^{RC}_{\omega,g}$. \textbf{B}. Comparison of $\mathcal{L}_{\omega,g}$ with the lower bounds $\mathcal{B}_{\omega,g}$ and $\mathcal{Q}_{\omega,g}$  (see the main text) at $\omega=3$ with $I^{RC}_{\eta,\omega}$ and the upper bound $\mathcal{U}_{\omega,\eta}$ as reference.}\label{fig:Fig2}
\end{figure}
\section*{Thermal amplification}
A thermal amplification channel, $\mathcal{E}_{g,\omega}$, is defined by the parameters $g>1$, and $\omega=2 n_{th} + 1 >1 $. The input-output relations of this channel are:
\begin{align}
	\bf{x}\to g\bf{x} && \bf{V}\to g\bf{V}+ (g-1)\omega \mathds{1},
\end{align}
Eve' optimal attack is the same of the previous section, an entanglement cloner. In the case of amplification however, the best performance is obtained in direct reconciliation, and the coherent information of the channel is:
\begin{equation}
	I^{C}_{g,\omega}= -\log_2\left(\frac{g}{g-1}\right)-h(\omega),
\end{equation}
the calculation of the $\Delta$ term yelds:
\begin{equation}
\Delta^G_{g,\omega}=\max_{\gamma>1} h(\lambda_+^{(E)})+h(\lambda_-^{(E)})-h\left(\gamma g-\omega+g\omega\right) \label{eq:deltaLoss g},
\end{equation}
Where once again we report the symplectic eigenvalues in Appendix.

In Fig.(\ref{fig:Fig2}) we report the same plots of the previous section. Namely, in panel \textbf{A} we plot the security threshold $\omega_{th}(g)$ of $\mathcal{L}^G_{g,\omega}$ and $I^{C}_{g,\omega}$. In panel \textbf{B} we compare $\mathcal{L}^G_{g,\omega}$ at a fixed value of $\omega$ to the best known bounds, taken from Ref.\cite{Wang_2019}, denoted $\mathcal{B}_{g,\omega}$, and Ref.\cite{Mele_2025}, denoted $\mathcal{Q}_{g,\omega}$, and with the upper bound on the capacity $\mathcal{U}=	I^{C}_{g,\omega}= -\log_2\left(\frac{g^{n_{th}}}{g-1}\right)-h(\omega)$. The results for thermal amplification are similar to the ones for thermal loss, with $\mathcal{L}^G_{g,\omega}$ saturated by $\mathcal{B}_{g,\omega}$, thus showing the optimality of the protocol in~\cite{Wang_2019}, within the restriction of our analysis, and $\mathcal{Q}_{g,\omega}$ being a small improvement for higher values of $g$.
\begin{figure}[]	\includegraphics[width=0.95\columnwidth]{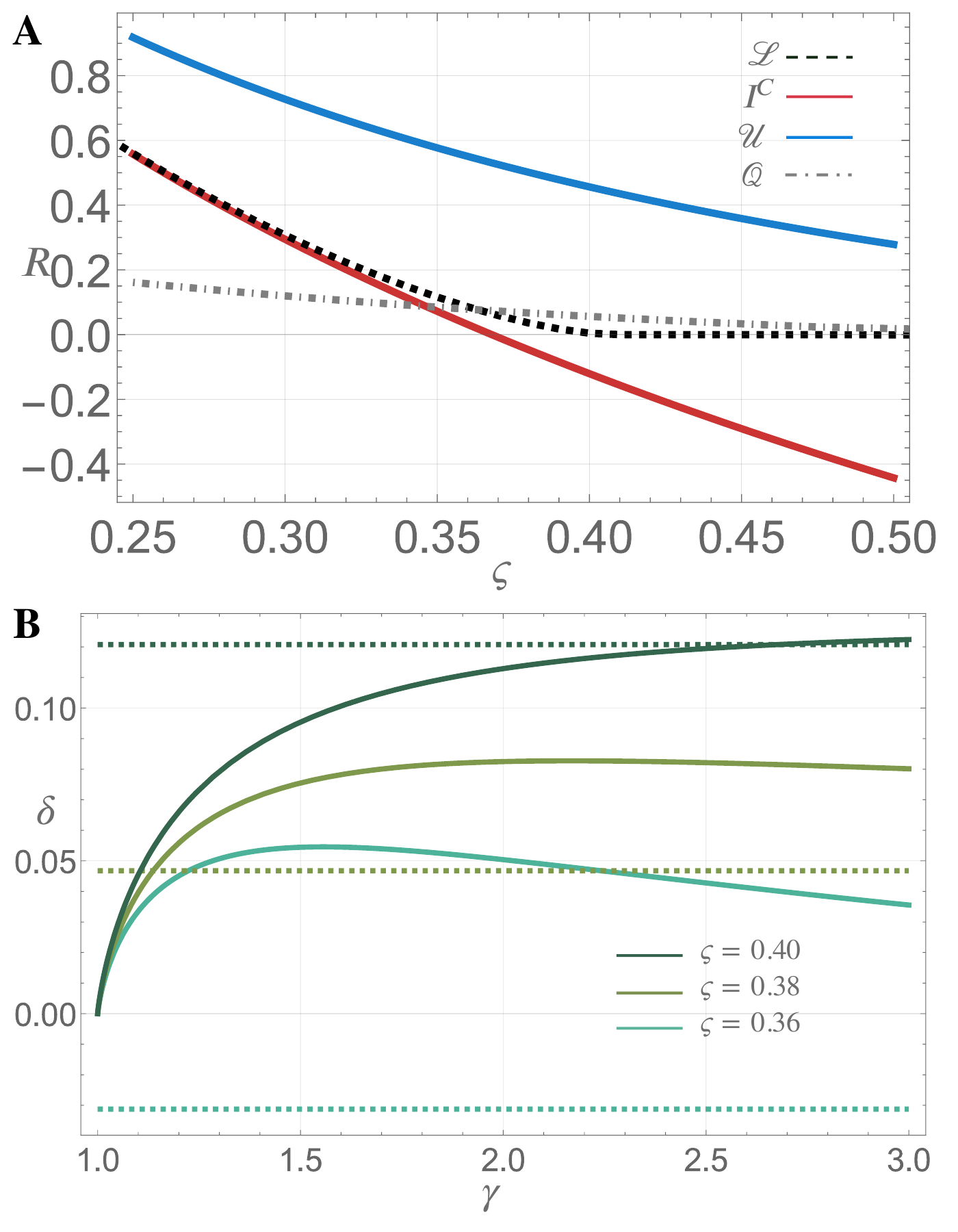}
  \caption{\textit{Added Noise.} \textbf{A}. Comparison of the bound $\mathcal{L}_\zeta^G$ with $I^{C}_{\zeta}$ and $\mathcal{Q}_\zeta$. The upper bound $\mathcal{U}_{\zeta}$ is also reported for reference. \textbf{B}. $\delta(\gamma)$ (see main text) is plotted for parameters $\bar{\zeta}=0.36,0.38,0.40$. The value of $-I^{C}_{\zeta}$ is reported for reference in dashed lines.}\label{fig:Fig3}
\end{figure}
\section*{Added noise}
An added noise channel, $\mathcal{E}_\zeta$, adds $\zeta>0$ thermal photons, representing classical noise, to the input signal. The input is transformed according to 	$\mathbf{x}\to\mathbf{x} + 2\zeta$ and $\mathbf{V}\to \mathbf{V}+ 2\zeta \mathds{1}$. The best lower bound for the private capacity of the channel is given by the coherent information~\cite{Pirandola_2009}:
\begin{equation}
I^{C}_{\zeta}= -\frac{1}{\log(2)}-\log_2\left(\zeta\right).
\end{equation}
We point out that for this channel the same lower bound is obtained by the reverse coherent information, $I^{RC}_{\zeta}=I^{C}_{\zeta}$ since they are computed in the limit $\mu\to\infty$. The quantity $\Delta^G_\zeta$ can then be computed either in direct of reverse reconciliation, but both configurations yield the same result.
For this channel the best collective attack is an universal cloner~\cite{Cerf_2000, Pirandola_2008}, different from the  entanglement cloner of previous section, as detailed in Appendix.
For the channel $\mathcal{E}_\zeta$ we compute:
\begin{equation}
\Delta^G_{g,\omega}=\max_{\gamma>1} h(\lambda_+^{(E)})+h(\lambda_-^{(E)})-h\left(2\zeta+\gamma \right), \label{eq:deltaNoise}
\end{equation}
where $\lambda_\pm^{(E)}$ are reported in Appendix.

In Fig.({\ref{fig:Fig3}}).\textbf{A} we plot $\mathcal{L}_\zeta^G$ as a function of $\zeta$ and compare it to $I^{C}_{\zeta}$ and the lower bound of Ref.\cite{Mele_2025}, denoted as $\mathcal{Q}_{\zeta}$. The plot shows how the former is an improvement on the latter for lower values of $\zeta$. We also report the upper bound on the capacity, $\mathcal{U}=(\zeta-1)/\log(2)-\log_2\left(\zeta\right)$~\cite{Pirandola_2017}, as a reference. 

The improvement over the coherent information is further showed in Fig.({\ref{fig:Fig3}}).\textbf{B} where we plot the   the argument of Eq.(\ref{eq:deltaNoise}), $\delta(\gamma):=h(\lambda_+^{(E)})+h(\lambda_-^{(E)})-h\left(2\zeta+\gamma \right)$, for  fixed values $\bar{\zeta}=0.36,0.38,0.40$, as a function of the optimization parameter $\gamma\geq 1$. Since the improvement is significant only when $I^{C}_{\zeta}+\delta(\gamma,\zeta)>0$ we also report the value $-I^{C}_{\bar{\zeta}}$, in dashed lines for each value of $\bar{\zeta}$, providing a visualization of the range of parameters $\gamma$ for which the bound is effectively tighter. Our bound thus provides an improvement, albeit small, on any previously known lower bounds.

\section*{Conclusions}
We analyzed a general protocol in which private communication between two parties is performed by locally measuring a bipartite state shared via a quantum channel $\mathcal{E}$, aided by one way classical communication. We provided a lower bound on the achievable secret key rate of this scheme, that in turns gives a lower bound on the secret key capacity of the channel. We focused our analysis on phase insensitive Gaussian channels and restricted the measurements to single mode Gaussian ones. Our analysis finds a relatively simple formula for a lower bound of the secret key capacity involving an optimization over a single real parameter. This lower bound shows the optimality, under the aforementioned conditions, of previously proposed protocols for the thermal-loss and thermal amplification channels and further gives an improved lower bound for the additive noise channel.

The analysis presented here can be expanded to other channels beside phase-insensitive ones, and further enhanced by dropping the restriction of Gaussian measurements. Another promising development would be to consider a similar scheme that allows multi-mode measurements.

\section*{Acknowledgments}
UKRI supported this work through the Integrated Quantum Networks (IQN) Research Hub (EPSRC, Grant No. EP/Z533208/1).
G.O.~and L.B.~ acknowledge financial support from:  
  PNRR Ministero Universit\`a e Ricerca Project No. PE0000023-NQSTI, funded by European Union-Next-Generation EU (G.O.,~L.B.), 
  the European Union’s Horizon Europe research and innovation program under EPIQUE Project GA No. 101135288.

\bibliography{bib}

\appendix
\section*{Appendix}
\subsection*{Optimization over Gaussian measurements}
In the protocol of the main text we denote the party performing the measurement $X$ and the receiving end of the communication $Y$, while the eavesdropper is denoted as $E$. According to Eq.(\ref{eq:deltaG}) the quantity $\Delta_G$ is fully determined by the the conditioned entropy terms $S(\rho_{Y/E|X})$. Since our protocol is fully Gaussian, the states $\rho_{Y/E|X}$ will be Gaussian as well and $S(\rho_{Y/E|X})$, in turn, fully determined by the CM $\mathbf{V}_{Y/E|X}$. Let us denote the terms of the bipartite CM $\mathbf{V}_{(Y/E )X}$ after the channel $\mathcal{E}$ as:
\begin{equation}
	\mathbf{V}_{(Y/E)X}=\begin{pmatrix}
		\mathbf{Y/E}& \mathbf{C}_{Y/E } \\
		\mathbf{C}_{Y/E}^T & \mathbf{X}
	\end{pmatrix}.
\end{equation}
After a Gaussian measurement on $X$ the conditioned local CM is:
\begin{equation}
\mathbf{V}_{D|X}=\mathbf{D}-\mathbf{C}(\mathbf{X}+\mathbf{V}_0)^{-1}\mathbf{C}^T, \label{eq:cstate}
\end{equation} 
where $D=Y/E$, $\mathbf{C}=\mathbf{C}_{Y/E }$ and $\mathbf{V}_0$ is the CM of a zero mean Gaussian state determining the measurement performed. As stated in the main text we can decompose $\mathbf{V}_0$ as:
\begin{equation}
	\mathbf{V}_0(\gamma,r,\theta)=\gamma\mathbf{R}(\theta)\mathbf{S}(2r)\mathbf{R}(\theta)^T,
\end{equation}
Using this decomposition the measurement $\Pi^G_X$ is determined by the parameters $(\gamma,r,\theta)$. Let us denote the argument of the maximization of $\Delta_G$ as $\delta(\gamma,r,\theta):=S(\rho_{E|X})-S(\rho_{Y|X})$.

To derive Eq.(\ref{eq:deltaPI}) we start by showing that $\delta$ does not depend on $\theta$. Consider the following equality:
\begin{align}
&\mathbf{C}(\mathbf{D}+\gamma\mathbf{R}(\theta)\mathbf{S}(2r)\mathbf{R}(\theta)^T )^{-1}\mathbf{C}^T= \nonumber \\
=&\mathbf{C}\mathbf{R}(\theta)\left(\mathbf{R}(\theta)^T \mathbf{X}\mathbf{R}(\theta) +\gamma\mathbf{S}(2r)\right)^{-1}\mathbf{R}(\theta)^T\mathbf{C}^T, \label{eq:equiv}
\end{align}
where we used the orthogonality of $\mathbf{R}(\theta)$, $\mathbf{R}(\theta)\mathbf{R}(\theta)^T=\mathds{1}$.
In the protocol analyzed, the initial states are fixed to $\Phi_A$ and $\Phi_E$ so after a phase-insensitive channel $\mathbf{R}(\theta)^T \mathbf{D}\mathbf{R}(\theta)= \mathbf{D}$. For the same reason $\mathbf{C}_{Y}\propto \mathbf{Z}=\text{diag}(1,-1)$, so that $\mathbf{R}(\theta)\mathbf{C}_{Y}\mathbf{R}(\theta)=\mathbf{C}_{Y}$. We have then:
\begin{equation}
\mathbf{R}(\theta)\mathbf{V}_{Y|X}\mathbf{R}(\theta)^T=\mathbf{Y}-\mathbf{C}_{Y}(\mathbf{X}+\gamma\mathbf{S}(2r))^{-1}\mathbf{C}_{Y}^T.
\end{equation}
In other words the parameter $\theta$ determines an unitary local rotation on the conditioned CMs $\mathbf{V}_{Y|X}$ and does not alter its entropy.
For system $E$ we note that $	\mathbf{C}_{E}=\begin{pmatrix}
		\mathbf{C}_1 \\
		\mathbf{C}_2
	\end{pmatrix} $ where the composing blocks are proportional to either $\mathbf{Z}$ or $\mathds{1}$. This means that it exist a local unitary, $\mathbf{R}_2(\theta_1,\theta_2):=\mathbf{R}(\theta_1)\oplus\mathbf{R}(\theta_2)$, on $E$ such that: 
	\begin{equation}
\mathbf{R}_2(\theta_1,\theta_2)\mathbf{V}_{E|X}\mathbf{R}_2(\theta_1,\theta_2)^T=\mathbf{E}-\mathbf{C}_{E}(\mathbf{X}+\gamma\mathbf{S}(2r))^{-1}\mathbf{C}_{E}^T,
\end{equation}
where $\theta_1,\theta_2$ are either $\theta$ or $-\theta$. This means that the conditional entropy on $E$ does not depend on $\theta$ as well.
Consequently, for phase-insensitive channels, $\delta(\gamma,r,\theta)$ does not depend on $\theta$,   $\delta(\gamma,r,\theta)= \delta(\gamma,r)$.

On the other hand the dependence of $\delta(\gamma,r)$ on the squeezing parameter $r$ is not trivial. However, for all the channels considered  we can directly compute the derivatives $\partial_r \delta(\gamma,r)|_{r=0}=0$ and $\partial^2_r \delta(\gamma,r)|_{r=0}<0$, showing that a maximum is achieved at $r=0$ and numerically check that the maximum is indeed global. Thus we have:
\begin{equation}
\Delta^G=\max_{\gamma\geq1} \delta(\gamma,0),
\end{equation}
that is Eq.(\ref{eq:deltaPI}) of the main text.
\subsection*{Collective attacks}
\begin{figure}[]	\includegraphics[width=0.8\columnwidth]{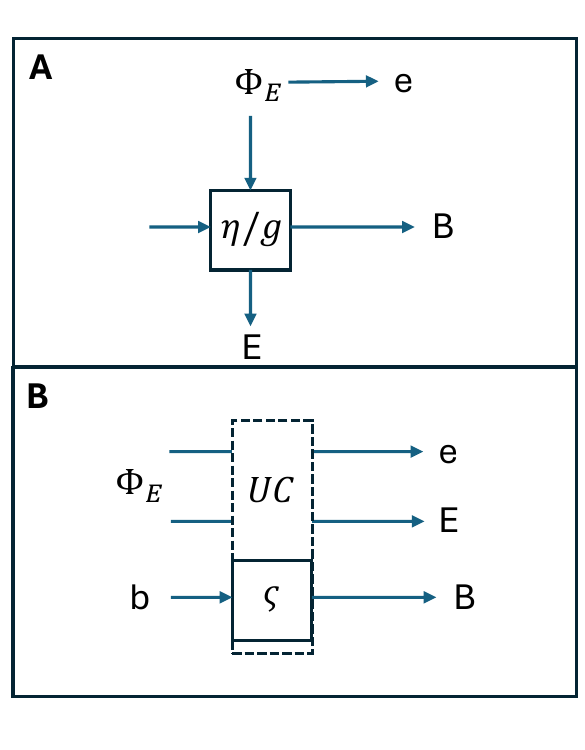}
	\caption{\textit{Dilation of the channels.}\textbf{A}. Thermal loss and thermal amplification are dilated by a beam splitter/thermal amplifier with parameter $\eta/g$. The dilation consists of a two-mode interaction between $B$ and $E$, while $e$ is stored for later measurement. \textbf{B}. The dilation of an added noise channel is a universal cloner, a three-mode interaction involving $e,E$ and $B$, described by the unitary in Eq.(\ref{eq:UC}).}\label{fig:scheme2}
\end{figure}
For the sake of security analysis, a quantum channel can be interpreted as the effect of an eavesdropper's, Eve, attack on the transmission. This is done by considering a dilation of the channel to a larger space with the assumption that this larger space is under Eve's control. We consider collective attacks meaning that Eve' state is stored in a quantum memory for the duration of the communication and global measurements are performed at the end of the protocol. In this scenario Eve' recovered information is upper bounded by the conditioned $\chi-$quantity, $\chi_{E|X}$, to the classical information $X$ as discussed in the main text. Collective attacks on Gaussian channel are fully characterized in terms of the Stinespring dilation where Eve's input state is a TMSV state of parameter $\omega$~\cite{Pirandola_2008}. 

For the thermal loss and thermal amplification channels, $\mathcal{E}_{\omega,\eta/g}$, this dilation is similarly constituted of an entanglement cloner~\cite{Pirandola_2008}. This consist in a two mode interaction, a beam splitter/thermal amplifier with transmissivity/amplification, $0\leq\eta\leq 1$ or $ g > 1$ respectivelly. As reported in Fig.(\ref{fig:scheme2}.\textbf{A}) one mode of Eve TMSV state enters the second port of the beam splitter/amplifier while the other mode is kept for joint measurements.

For the added noise channel, $\mathcal{E}_{\zeta}$, the optimal collective attack is an universal cloner~\cite{Cerf_2000}. It consist in an irreducible three mode channel composed by a sequence of cv-CNOT gates~\cite{Bartlett_2002}. A continuous variable CNOT in the phase space is an unitary operation acting on the quadratures of two modes, $\{\hat{x}_j,\hat{p}_j\}$, as:
\begin{equation}
U_{\text{CNOT}}=e^{-i c \hat{x}_1 \hat{p}_2},
\end{equation}
with the real parameter $c$ determining the strength of the interaction.
Referring to the scheme in Fig.(\ref{fig:scheme2}.\textbf{B}) for the labeling of the modes, the universal cloner acts as~\cite{Cerf_2000}:
\begin{equation}
U_{\text{UC}}=e^{-i(\hat{x}_e-\hat{x}_E) \hat{p}_B}e^{-i\hat{x}_B(\hat{p}_e-\hat{p}_E)}. \label{eq:UC}
\end{equation}
The CM of the full system evolves according to $\mathbf{V}_{eEB}\to\mathbf{L}_{UC}\mathbf{V}_{eEB}\mathbf{L}_{UC}^T$, where $\mathbf{L}_{UC}$ is the symplectic matrix associated with $U_{\text{UC}}$:
\begin{equation}
\mathbf{L}_{UC}=\begin{pmatrix}
1 & 0 & 0 & 0 & 2 & 0 \\
0 & 5 & 0 & 4 & 0 & -2 \\
0 & 0 & 1 & 0 & 2 & 0 \\
0 & -4 & 0 & -3 & 0 & 2 \\
2 & 0 & -2 & 0 & 1 & 0 \\
0 & -2 & 0 & -2 & 0 & 1
\end{pmatrix}.
\end{equation}
The local action on $B$ of this transformation is:
\begin{equation}
\mathbf{B}\to\mathbf{B}+2\left(4\omega - 4\sqrt{\omega^2-1}\right)\mathds{1},
\end{equation}
that is an added noise channel with parameter $\zeta=4\omega - 4\sqrt{ \omega^2-1}$. 
\subsection*{Symplectic eigenvalues}
In this section we report the symplectic eigenvalues of Eve's covariance matrix, $\mathbf{V}_E$, used in the main text analysis for the thermal-loss, thermal amplification and added noise channels.
\subsubsection*{Thermal Loss}
\begin{align*}
  \lambda_{\pm}^{(E)} &= \frac{\sqrt{A \pm \sqrt{B}}}{\sqrt{2}\eta}, \\
  A &= (\eta-1)^2 (\gamma - \omega)^2 + 2\eta, \\
  B &= (\eta-1)^2 (\gamma + \omega)^2 \times \\ & ~~~~~~~
  \left( (\eta-1)^2 (\gamma - \omega)^2 - 4\gamma\omega(\eta - 1) + 4\eta \right).
\end{align*}
\subsubsection*{Thermal Amplification}
\begin{align*}
  \lambda_{\pm}^{(E)} &= \frac{\sqrt{A \pm \sqrt{B}}}{\sqrt{2}}, \\
  A &= (g - 1)^2 (\gamma^2 + 2g\gamma\omega + \omega^2) + 2g, \\
  B &= (g - 1)^2 (\gamma + \omega)^2 \big[ \gamma^2(g - 1)^2 - \\ &~~~~~~
  - 2\gamma(g^2 - 1)\omega + (g - 1)^2\omega^2 + 4g \big].
\end{align*}
\subsubsection*{Added Noise}
\begin{align*}
\lambda_{\pm}^{(E)} = \sqrt{1 + \frac{1}{2} \zeta \left( \zeta + 2\gamma \pm \sqrt{4 + \zeta^2 + 4\zeta\gamma} \right)}.
\end{align*}
\subsection*{Secret Key Rate}
We lower bound the optimal secret key rate in the asymptotic regime of a large number of uses of the channel. 

Following the notation of the main text we label the party performing the measurement as $X$, and the other party as $Y$. Let us fix the measurement $\Pi_X$, yielding the classical random variable $\mathbf{x}$ on system $X$. Communication is achieved by subsequent measurements on $Y$, aided by one way classical communication.
For the classical-quantum system $XY$ the maximum common randomness that can be extracted with unlimited one way classical communication is given by the classical-quantum Slepian-Wolf result \cite{Devetak_2003} (see also Lemma 4 of \cite{Devetak_2004}), stating that $S(\textbf{x}|Y)$ bits of public communication are needed from $X$ to $Y$, for the latter to reproduce $\textbf{x}$ (in the asymptotic limit of many copies). This yields a raw key rate:
\begin{equation}
R_{XY}=H(\textbf{x})-S(\textbf{x}|Y)=I(\textbf{x}:Y),
\end{equation}  
where $S(\textbf{x}|Y)$ are the bits of public communication and thus subtracted from the key, and $I(\mathbf{x}:Y)$ is the mutual information of the classical quantum system $XA$, equal to the Holevo quantity $\chi_{Y}^{\Pi_X}$, i.e. $I(\mathbf{x}:Y)=\chi_{Y}^{\Pi_X}$. 
After a collective attack by Eve the optimal asymptotic secret key rate, $R^{\Pi_X}$, is lower bounded using the above raw key rate and the classical Csiszar-Korner theorem by the quantity \cite{Csiszar_2011,Pirandola_2020}:
\begin{equation}
R^{\Pi_X} \geq R_{XY}-I(\mathbf{x}:E), \label{eq:rate1}
\end{equation}
where $I(\mathbf{x}:E)=\chi_{E}^{\Pi_X}$ serves as an upper limit on the information extracted by Eve regardless of measurement and post-processing. 
Maximizing the rate in Eq.(\ref{eq:rate1}) over $\Pi_x$, in the limit of $\mu \to \infty$, gives our lower bound in Eq.(\ref{eq:rate}).
\end{document}